\documentstyle[seceq,preprint]{jpsj}

\def\runtitle{Anisotropy of Mott-Hubbard gap transitions due to 
spin and orbital ordering in LaVO$_{3}$ and YVO$_{3}$}
\def\runauthor{Shigeki {\sc Miyasaka}, Yoichi {\sc Okimoto} 
and Yoshinori {\sc Tokura}}

\title
{
Anisotropy of Mott-Hubbard Gap Transitions due to 
Spin and Orbital Ordering in LaVO$_{3}$ and YVO$_{3}$
}

\author
{ 
Shigeki {\sc Miyasaka}$^{1,2}$, 
Yoichi {\sc Okimoto}$^{1}$ and 
Yoshinori {\sc Tokura}$^{1,2,3}$
}

\inst
{
$^1$Correlated Electron Research Center (CERC), 
National Institute of Advanced Industrial Science and Technology 
(AIST), Tsukuba 305-8562

$^2$Department of Applied Physics, University of Tokyo, Tokyo 
113-8656

$^3$Spin Superstructure Project, ERATO, Japan Science and Technology 
Corporation (JST), Tsukuba 305-8562
}

\recdate
{
\today
}

\abst
{
Anisotropic optical spectra coupled with antiferromagnetic spin ordering (SO) 
and orbital ordering (OO) have been investigated for single crystals of 
LaVO$_{3}$ and YVO$_{3}$. 
The orbital-dependent Mott-Hubbard gap transitions are observed around 2 eV. 
The transitions composed of the two peaks show distinct anisotropy and 
selection rules, reflecting the respective SO and OO patterns. 
The temperature dependence of the anisotropic transitions clearly indicates 
the SO/OO correlation and the evolution of the order parameters.
}

\kword
{
orbital-dependent Mott-Hubbard gap transitions, spin and orbital ordering, 
LaVO$_{3}$, YVO$_{3}$
}

\begin{document}
\sloppy
\maketitle

As a consequence of strong electron correlation, 
tight coupling among spin, charge and orbital degrees of freedom 
produces a variety of intriguing physical properties 
in 3$d$ transition-metal oxides~\cite{Mott,Imada}. 
An example of such a property is the colossal magnetoresistance and 
related phenomena for the hole-doped perovskite-type manganites, 
in which spin ordering (SO) and orbital ordering (OO) govern 
the charge dynamics and electronic structure~\cite{Tokura}. 
Vanadates, as well as manganites, are regarded 
as one of the prototypical systems that show spin-charge-orbital coupled 
phenomena~\cite{Castellani,Rice,Bao,Paolasini,Khaliullin,Noguchi,Miyasaka}. 
To investigate charge dynamics coupled with the spin and 
orbital correlations, we have revisited classic vanadium oxides, 
perovskite-type LaVO$_{3}$ and YVO$_{3}$.

LaVO$_{3}$ and YVO$_{3}$ are prototypical Mott-Hubbard type insulators 
with the V electron configuration of 3$d^{2}$, 
and have a $Pbnm$ orthorhombic unit cell with 
$a \approx b \approx c/ \sqrt{2}$ at room temperature~\cite{Imada}. 
These compounds undergo a magnetic transition to 
an antiferromagnetic (AF) state as well as 
a structural phase transition due to the OO 
as temperature is decreased. 
The schematic structures of the SO and OO in both compounds 
are shown in Figs. 1(a) and 1(b). 
In LaVO$_{3}$, the magnetic transition occurs at 143 K, 
and the ordered spin configuration is $C$-type with an AF coupling 
in the $ab$-plane and a ferromagnetic one 
along the $c$-axis~\cite{Zubkov}. 
This compound undergoes another phase transition, 
a first-order structural transition 
from an orthorhombic to a monoclinic form at 141 K~\cite{Bordet}. 
Sawada $et$ $al.$ have theoretically demonstrated that LaVO$_{3}$ 
in the monoclinic phase has an OO 
with an alternative $d_{xy}^{1}d_{yz}^{1}/d_{xy}^{1}d_{zx}^{1}$ 
electron configuration (Fig. 1(a))~\cite{Sawada}. 
Here, we call this OO $G$-type by analogy to spin ordering. 
In YVO$_{3}$, which is attracting interest 
because of a temperature-induced magnetization 
reversal phenomenon~\cite{Ren}, synchrotron $X$-ray diffraction, 
neutron and resonant $X$-ray scattering studies 
have confirmed the existence of two AF spin- and 
orbital-ordered states~\cite{Blake,Noguchi,Kawano}. 
Between the first and second N$\acute{\rm e}$el temperatures 
($T_{N1}$=115 K$>T>T_{N2}$=71 K), 
this compound shows the $C$-type SO, and hence has been supposed to 
have SO and OO identical to those of the ground-state 
in LaVO$_3$(Fig. 1(a)). 
Below $T_{N2}$, a distortion of octahedral VO$_{6}$ 
occurs due to the Jahn-Teller effect, and the SO pattern changes 
into the $G$-type with the antiferromagnetically arranged V$^{3+}$ spins 
in all three directions, while the OO turns into the $C$-type 
with the alternative $d_{xy}^{1}d_{yz}^{1}/d_{xy}^{1}d_{zx}^{1}$ 
electron configuration in the $ab$-plane 
and the identical one along the $c$-axis (Fig. 1(b)). 
The structures of the SO and OO in LaVO$_3$ and YVO$_3$ 
are thus highly anisotropic in spite of the almost cubic lattice structure. 
The SO and OO are expected to cause appreciable anisotropy 
in the electronic structure and charge dynamics 
through the spin-charge-orbital correlations. 
In this paper, we have investigated polarized optical 
conductivity spectra (electric field, {\bf E}$\parallel$$c$-axis 
({\bf E}$\parallel$$c$) 
and {\bf E}$\perp$$c$-axis ({\bf E}$\perp$$c$)) 
in the perovskite-type LaVO$_3$ and YVO$_3$ single crystals, 
to clarify the anisotropic charge dynamics 
coupled with the SO and OO, as well as to probe 
the spin-orbital correlations.

Single crystals of LaVO$_{3}$ and YVO$_{3}$ were grown 
by a floating zone method in an atmosphere of Ar, 
as reported elsewhere~\cite{Noguchi,Miyasaka}. 
The melt-grown crystal of YVO$_3$ was confirmed to be 
a single-domain crystal by Laue reflection and single-crystal 
$X$-ray diffraction with a four-circle diffractometer. 
The results of the $X$-ray analyses indicated 
that the crystal of LaVO$_3$ was also detwinned 
in the $ac$- and $bc$-planes, 
although we did not confirm whether it was detwinned 
in the $ab$-plane because of the nearly identical 
lattice parameters for the $a$- and $b$-axes 
at room temperature~\cite{Bordet}. 
In the present experiment, this form of LaVO$_3$ crystal 
was sufficient for the optical reflectivity measurements, 
since the structures of the SO and OO can be viewed 
almost isotropic in the $ab$-plane. 
The crystals of both compounds with a (100) or (010) surface were 
polished to optical flatness with alumina powder. 
To remove the mechanical stress induced by surface polishing, 
we annealed the crystals at 1000 $^{\circ }$C for 24 h 
in a flow of forming gas of Ar/H$_{2}$ (93/7$\%$). 
Reflectivity measurements were carried out 
on the (100) or (010) surface between 0.06 and 40 eV. 
We used a Fourier-transform interferometer for the 0.06-0.8 eV region 
and grating spectrometers for the higher-energy region (0.6-40 eV). 
Measurements with varying temperature were performed 
between 0.06 and 5 eV. 
The synchrotron radiation source (UV-SOR) 
at the Institute for Molecular Science 
was utilized for spectroscopy above 5 eV at room temperature. 
Optical conductivity ($\sigma$($\omega$)) spectra 
at various temperatures were obtained 
by Kramers-Kronig analysis of the respective reflectivity spectra 
combined with the room-temperature spectrum above 5 eV. 
For the analysis, we assumed a constant reflectivity below 0.06 eV 
and a $\omega^{-4}$-type extrapolation above 40 eV.

Figure 2 presents the $\sigma$($\omega$) spectra 
for {\bf E}$\parallel$$c$ and {\bf E}$\perp$$c$ 
in LaVO$_{3}$ and YVO$_{3}$ at various temperatures~\cite{phonon}. 
In both vanadates, the {\bf E}$\parallel$$c$ $\sigma$($\omega$) spectra 
show large temperature ($T$) -dependence, 
while the spectra for {\bf E}$\perp$$c$ are almost independent of $T$. 
In LaVO$_3$, the spectral weight of the $\sigma$($\omega$) for 
{\bf E}$\parallel$$c$ 
in a higher-energy region is rapidly transferred 
across the isosbestic (equal absorption) point at 2.9 eV 
to a lower-energy region 
with the decrease of $T$ around the magnetic and 
structural phase transition temperatures ($T_N$ and $T_S$). 
As a result of the spectral weight transfer, 
the peak intensity around 2 eV is markedly enhanced in the $C$-type 
spin- and $G$-type orbital-ordered state. 
In YVO$_3$, on the other hand, 
the spectral weight of the $c$-axis polarized $\sigma$($\omega$) 
is transferred from the region above 3.4 eV to a lower-energy region, 
and the peak intensity around 2 eV significantly increases 
with decreasing $T$ between $T_{N1}$ and $T_{N2}$. 
Moreover, the $\sigma$($\omega$) for {\bf E}$\parallel$$c$ shows 
further accumulation of the spectral weight below 3.4 eV 
in the $G$-type spin- and $C$-type orbital-ordered state ($T<T_{N2}$). 
In both compounds, the peak intensity around 2 eV 
in the $\sigma$($\omega$) is markedly enhanced only for 
{\bf E}$\parallel$$c$ 
with decreasing $T$. 
The 2 eV band in the $\sigma$($\omega$) spectrum 
has been assigned to the Mott-Hubbard gap 
transition~\cite{Arima,Inaba,Kasuya,Maiti,CT}. 
In the AF phase, however, the peak around 2 eV in the 
{\bf E}$\parallel$$c$ spectra 
shows a distinct shape, 
which has seldom been observed in other three-dimensional 
Mott-Hubbard insulators~\cite{Arima,Thomas}. 
Such an anisotropic and sharp spectral feature 
for the Mott-Hubbard gap transition reflects 
the anisotropic electronic structure due to the SO and OO 
with one-dimensional character, as argued in the following.

As seen in Fig. 2, the 2 eV band 
in the $\sigma$($\omega$) for {\bf E}$\parallel$$c$ is composed of two peaks 
at 1.77 eV and 2.41 eV in LaVO$_3$, and at 1.88 eV and 2.37 eV in YVO$_3$. 
The {\bf E}$\parallel$$c$ $\sigma$($\omega$) spectra around 2 eV can be fitted 
with the Lorentz model, 
expressed by the following formula:

\begin{equation}
 \sigma(\omega) = \sum_{i=1}^{3} f_{i} (\omega)
= \sum_{i=1}^{3} \frac{S_{i} \omega_{i}^{2} \gamma_{i} \omega^{2}}
{(\omega_{i}^{2} - \omega^{2})^{2}+\gamma_{i}^{2} \omega^{2}}.
\label{eq:Lorentz}
\end{equation}
Here, the simulated spectra expressed by the functions 
($f_{i}(\omega)$, $i=1-3$) correspond to the lower-lying peak 1, 
the higher-lying peak 2 and the peak for the charge-transfer gap 
transition located above 4 eV. 
$\gamma_{i}$, $\omega_{i}$ and $S_{i}$ are width, 
excitation frequency and oscillator strength, respectively. 
The insets of Figs. 3 and 4 exemplify the results of the fitting 
at 10 K for the both vanadates. 
At the ground state (10 K), peak 1 bears 
much larger (smaller) intensity than peak 2 
in LaVO$_3$ (YVO$_3$), as clearly seen in this spectral decomposition. 
In both vanadates, an electron can hop 
only between the $d_{yz}$ orbitals or the $d_{zx}$ ones 
on the neighboring V sites along the $c$-axis 
through the $\pi$-bonding with the O 2$p_y$ or 2$p_x$ state. 
The $d_{yz}$-$d_{yz}$ transition energy should be 
nearly equal to the $d_{zx}$-$d_{zx}$ one. 
In Figs. 1(c) and 1(d), we show the initial and possible final states 
of the $t_{2g}$ orbitals on the neighboring V sites 
along the $c$-axis in the optical Mott-Hubbard gap transitions 
in the respective cases of the SO and OO. 
In both cases, the $d_{zx}$ electron can be 
transferred to the $d_{zx}$ orbital on the nearest-neighbor V site 
along the $c$-axis. 
The energy of the $d_{zx}$-$d_{zx}$ excitation in the $C$-type SO and 
$G$-type OO is smaller than that in the $G$-type SO and $C$-type OO 
by an energy of the effective Hund's-rule coupling. 
Therefore, in the case of full polarization of the spin and orbital, 
the $\sigma$($\omega$) spectrum for {\bf E}$\parallel$$c$ would exhibit 
only one peak 
at $U'-K$ for the $C$-type SO and $G$-type OO, 
or otherwise at $U'+K$ for the $G$-type SO and $C$-type OO, 
where $U'$ and $K$ denote the energy of effective 
on-site Coulomb repulsion and Hund's-rule coupling, respectively. 
Such orbital-dependent Mott-Hubbard gap transitions 
and their selection rules in the spin- and orbital-ordered state 
can explain heavily the unbalanced two-peak features 
in the respective ground-state ($e.g.$, 10K) spectra 
of LaVO$_3$ and YVO$_3$.

Neutron scattering studies have revealed 
that the magnetic moments are approximately 1.3 $\mu _{B}$ for LaVO$_3$, 
and 1.6 $\mu _{B}$ below $T_{N2}$ and 1.0 $\mu _{B}$ between $T_{N1}$ 
and $T_{N2}$ for YVO$_3$, while the pure spin value is 
2 $\mu _{B}$ for $S$=1~\cite{Zubkov,Kawano}. 
The excitation with the energy of $U'+K$ may be optically allowed 
by the imperfect polarization of the spin in the $C$-type SO 
even under the complete polarization of the orbital in the $G$-type OO. 
Moreover, the transition with the energy of $U'-K$ becomes weakly 
allowed in the $G$-type SO and $C$-type OO, 
if both orbital and spin show imperfect polarization. 
Therefore, the imperfect polarization of the spin and 
perhaps that of the orbital as well may relax the above-mentioned 
selection rule 
and produce the two-peak feature in the 
{\bf E}$\parallel$$c$ $\sigma$($\omega$) 
spectra. 
The orthorhombic distortion in the actual crystal structure 
and/or the next-nearest-neighbor electron hopping 
should further relax the selection rule, 
although the large difference in the respective peak intensities 
should remain appreciable as observed.
Thus assigning the two-peak feature 
to the orbital-dependent Mott-Hubbard transitions ($U'-K$ and $U'+K$), 
$U'$ and $K$ are estimated to be approximately 2.09 eV and 0.32 eV 
for LaVO$_3$, and 2.13 eV and 0.25 eV for YVO$_3$ 
on the basis of the localized electron picture. 
Incidentally, an electron can be transferred in the $ab$-plane 
not only between the $d_{xy}$ orbitals but also 
between the $d_{yz}$ or $d_{zx}$ ones on the neighboring V sites 
in both spin- and orbital-ordered states, 
and the energy of the excitations is $U'+K$. 
It is noted, however, that the $ab$-plane V-O bonds 
are longer than the out-of-plane 
($c$-axis) ones even at high temperatures. 
Below $T_S$ and $T_{N2}$, the V-O bonds are further 
stretched due to the occupied $d_{yz}$ or $d_{zx}$ orbital, 
and the octahedral VO$_6$ units tilt (rotate) 
in the $ab$-plane~\cite{Bordet,Kawano}. 
Consequently, the O 2$p$-V 3$d$ transfer interaction 
is suppressed in the spin- and orbital-ordered state. 
This explains why the peak intensity in the $\sigma$($\omega$) 
for {\bf E}$\perp$$c$ remains low in the AF phase, 
in contrast to the {\bf E}$\parallel$$c$ case.

Figure 3 shows the $T$ dependence of the spectral weight 
($N_{{\rm eff}, \; i}$) of each peak ($i=$ 1 or 2) 
in LaVO$_3$ for {\bf E}$\parallel$$c$ 
that is defined as 

\begin{equation}
 N_{{\rm eff}, \; i} = \frac{2m_{0}}{\pi e^{2}N} 
\int_{0}^{\infty} f_{i}(\omega) d \omega 
= \frac{m_{0} S_{i} \omega _{i} ^{2}}{e^{2}N} \; \; \; (i = 1,2).
\label{eq:Spectral weight}
\end{equation}
Here, $m_{0}$ is the free electron mass and  
$N$ the number of V atoms per unit volume.
The spectral weight of peak 1 rapidly increases 
with decreasing $T$ below $T_N$ and $T_S$, 
while that of peak 2 is almost independent of $T$. 
As a result, the spectral weight of peak 1 is four times 
as large as that of peak 2 at the lowest $T$. 
The distinct enhancement of the spectral weight of peak 1 
immediatly below $T_N$ and $T_S$ is consistent with the aforementioned 
SO/OO scenario and hence ensures the emergence of 
the $C$-type SO and $G$-type OO in LaVO$_3$. 
Peak 1 shows a slight accumulation of the spectral weight 
even above $T_N$ and $T_S$ with the decrease of $T$. 
The result implies the subsisting correlation of 
the $C$-type SO and $G$-type OO above $T_N$ and $T_S$. 
The correlation of the SO and OO is likely to cause 
the small anisotropy between the $\sigma$($\omega$) spectra 
for {\bf E}$\parallel$$c$ and {\bf E}$\perp$$c$ even around room 
temperature, as observed in Fig. 2(a).

In YVO$_3$ (Fig. 4), the spectral weight of peak 2 is much larger 
than that of peak 1 at temperatures below $T_{N2}$. 
This is quite consistent with the predicted low-$T$ 
electronic state of YVO$_3$, $i.e.$, the $G$-type SO and $C$-type OO. 
In the intermediate region, $T_{N2}<T<T_{N1}$, on the other hand, 
the $T$ dependent behavior of the respective spectral weights 
shown in Fig. 4 cannot be explained as straightforwardly 
as expected from the proposed spin-orbital order structure, $i.e.$, 
the $C$-type SO and $G$-type OO. 
The increase of the spectral weight for peak 1 below $T_{N1}$ 
is not inconsistent with the simple prediction. 
However, much larger increase of the intensity of peak 2 is 
observed below $T_{N1}$, and resultantly peak 2 bears 
a larger spectral weight than peak 1 in this intermediate $T$ region. 
The result suggests that the electronic state with the $C$-type SO 
realized in this intermediate $T$ region considerably differs from 
the ground state of LaVO$_{3}$ 
with a similar $C$-type SO. 
Since the relative intensity between peaks 1 and 2 is 
quite sensitive to the orbital order or correlation, 
the straightforward conclusion derived from the present observation 
is that even in the spin $C$-type 
state the $C$-type orbital correlation is present other 
(dominant rather) than the $G$-type correlation. 
As additional evidence of this, a recent neutron scattering study 
on the single-crystalline YVO$_3$~\cite{Keimer} has revealed 
that the SO at $T_{N2}<T<T_{N1}$ is not simple $C$-type 
but appreciably affected by the $G$-type canting 
with a canting angle of $\pm$10 deg. 
Nevertheless, the hypothesis of the dominant $C$-type orbital correlation 
at $T_{N2}<T<T_{N1}$ in YVO$_3$ is not consistent with the results of 
the recent synchrotron $X$-ray diffraction studies~\cite{Blake}, 
that clearly indicate the lattice modulation being consistent with 
the $G$-type OO in this $T$ region.

As a possible scenario that may solve the apparent discrepancy, we may 
consider a role of the large orbital fluctuation~\cite{Khaliullin}. 
Khaliullin $et$ $al.$~\cite{Khaliullin} have recently pointed out 
that the quantum fluctuation between the $d_{yz}$ and $d_{zx}$ 
orbitals (while $d_{xy}$ being occupied on every V site) 
is possibly quite large and tends to stabilize the ferromagnetic 
interaction along the $c$-axis as realized in the spin $C$-type state. 
Thus, the orbital state in the $C$-type SO of YVO$_3$ 
($T_{N2}<T<T_{N1}$) might be better described 
by the orbital singlet correlation rather than 
by the simple $G$-type orbital order, in particular along the $c$-axis. 
This might allow the evolution of the $G$-type magnetic correlation 
that was observed to mix in the spin $C$-type order~\cite{Keimer} 
and the resultant $C$-type orbital correlation 
along the $c$-axis as observed in the present spectroscopic analysis. 
However, the spectral shape and the spectral weight ratio immediately 
above $T_{N2}$ rather resemble those of the ground state, 
and such a significant feature remains puzzling and should be 
theoretically elucidated.

In summary, we have investigated the effect of the spin and orbital 
ordering (SO and OO) and their correlation on the optical gap 
transitions and their anisotropy. 
In LaVO$_3$ and YVO$_3$, the spectra for the {\bf E}$\parallel$$c$-axis 
show two peaks 
around 2 eV, which correspond to the orbital-dependent 
Mott-Hubbard gap transitions in the $C$-type SO and $G$-type OO, 
and those in the $G$-type SO and $C$-type OO, respectively. 
In the ground state of both vanadates, 
the spectral weight of the respective peaks and the anisotropy 
clearly reflect each type of the SO and OO.

We would like to thank N. Nagaosa, Y. Motome, Y. Tomioka, T. Arima 
and B. Keimer 
for helpful discussions. This work was supported in part by NEDO.


\clearpage

\begin{figure}
\caption{(a, b) Schematic structures of the $C$-type spin ordering (SO) 
and $G$-type orbital ordering (OO), 
and the $G$-type SO and $C$-type OO in the perovskite-type 
LaVO$_{3}$ and YVO$_{3}$. 
Open arrows indicate spins and gray and black lobes indicate 
occupied $d_{yz}$ and $d_{zx}$ orbitals 
on the vanadium ions, respectively. 
The commonly occupied $d_{xy}$ orbitals are displaced for clarity. 
(c, d) Initial states (upper pictures) and possible final ones 
(lower ones) of $t_{2g}$ orbitals on the neighboring V sites 
(V$_1$ and V$_2$) along the $c$-axis 
in the optical $d_{zx}$-$d_{zx}$ transition in each type of SO and OO.}
\label{fig:1}
\end{figure}

\begin{figure}
\caption{Optical conductivity spectra for {\bf E}$\parallel$$c$ and 
{\bf E}$\perp$$c$ 
(black lines) in single crystals of (a) LaVO$_{3}$ and (b) YVO$_{3}$. 
For LaVO$_{3}$, red and blue lines indicate the spectra for 
{\bf E}$\parallel$$c$ 
at temperatures (10-142 K) 
below the magnetic transition temperature ($T_{N}$), 
and at 150-293 K above $T_{N}$, respectively. 
For YVO$_{3}$, green, red and blue lines show the c-axis polarized spectra 
in the $G$-type spin- and $C$-type orbital-ordered state 
(at 10-67 K), in the $C$-type spin- and $G$-type 
orbital-ordered one (at 75-110 K) 
and in the paramagnetic one (at 120-293 K), respectively.}
\label{fig:2}
\end{figure}

\begin{figure}
\caption{Inset shows the fitting result with the 
Lorentz model for the $c$-axis polarized optical conductivity spectrum 
($\sigma$($\omega$)) at 10 K in LaVO$_{3}$. 
The lower- and higher-lying peaks are the Mott-Hubbard gap transitions, 
denoted ``peak 1" and ``peak 2", respectively. 
The solid black and red, blue, violet and green broken lines 
indicate the experimental and fitted curves for peak 1, 
peak 2, the charge-transfer gap transition and total components, respectively. 
The main panel presents the temperature dependence of the spectral weight 
for peak 1 (solid circles) and peak 2 (open circles). 
(See text for definition.) 
The spectral weight of peak 1 rapidly increases 
with decreasing temperature below the magnetic and 
structural phase transition temperature ($T_{N}$ and $T_{S}$), $i.e.$, 
in the $C$-type spin-ordered ($C$-type SO) and perhaps 
$G$-type orbital-ordered ($G$-type OO) state.}
\label{fig:3}
\end{figure}

\begin{figure}
\caption{Inset shows the fitting result with the 
Lorentz model for the $c$-axis polarized optical conductivity spectrum 
($\sigma$($\omega$)) at 10 K in YVO$_{3}$. 
The lower- and higher-lying peaks are the Mott-Hubbard gap transitions, 
denoted ``peak 1" and ``peak 2". 
The solid black and red, blue, violet and green broken lines 
indicate the experimental and fitted curves for peak 1, 
peak 2, the charge-transfer gap transition and total components, respectively. 
The main panel presents the temperature dependence of the spectral weight 
for peak 1 (solid circles) and peak 2 (open circles). 
(See text for definition.) 
The spectral weight of peak 2 is much larger than that of peak 1 
in the $G$-type spin- and $C$-type orbital-ordered state ($G$-SO and $C$-OO) 
below $T_{N2}$.}
\label{fig:4}
\end{figure}


\begin{thebibliography}{99}
\bibitem{phonon}Sharp spikes around 0.08 eV are due to optical-phonon modes.
\bibitem{CT}Above the Mott-Hubbard gap, a more intense optical transition is 
observed above 4 eV that is assigned to the charge-transfer-type 
transition from the occupied O 2$p$ state to the V 3$d$ 
upper Hubbard state~\cite{Arima,Kasuya}.
\end{thebibliography}
\end{document}